\documentclass[prl,twocolumn,showpacs,groupedaddress,superscriptaddress,amsmath,amssymb]{revtex4-2} 
 
\usepackage[utf8]{inputenc}
\usepackage[english]{babel}	

\usepackage{stfloats}

\usepackage{cmap}
\usepackage{bbm}
\usepackage{fontenc}

\usepackage{enumerate}
\usepackage{multirow}
\usepackage{float}
\usepackage{comment}
\usepackage{lineno}

\usepackage{amsmath,amsfonts,amssymb,mathtools}	
\usepackage{euscript} 
\usepackage{mathrsfs}
\usepackage{mathtext}

\usepackage{color}			
\usepackage{graphicx}
\usepackage[export]{adjustbox}
\DeclareGraphicsExtensions{.png,.jpg}
\graphicspath{}	
\usepackage[section]{placeins} 

\usepackage{physics}
\usepackage{braket}
\usepackage{xparse}	

\usepackage{slashed}
\usepackage{indentfirst} 
\usepackage{enumitem}

\usepackage[colorlinks=true, filecolor=blue, citecolor=blue,urlcolor=black]{hyperref}
\usepackage{orcidlink}
\usepackage{url}
\usepackage{cleveref}
\DeclareUnicodeCharacter{2212}{-}

\date{\today}

\begin{document}
\title{Exploration of the Muon $g-2$ and Light Dark Matter explanations in NA64 with the CERN SPS high energy muon beam}
\author{Yu.~M.~Andreev\orcidlink{0000-0002-7397-9665}}
\affiliation{Authors affiliated with an institute covered by a cooperation agreement with CERN}
\author{D.~Banerjee\orcidlink{0000-0003-0531-1679}}
\affiliation{CERN, European Organization for Nuclear Research, CH-1211 Geneva, Switzerland}
\author{B.~Banto Oberhauser\orcidlink{0009-0006-4795-1008}}
\affiliation{ETH Z\"urich, Institute for Particle Physics and Astrophysics, CH-8093 Z\"urich, Switzerland}
\author{J.~Bernhard\orcidlink{0000-0001-9256-971X}}
\affiliation{CERN, European Organization for Nuclear Research, CH-1211 Geneva, Switzerland}
\author{P.~Bisio\orcidlink{/0009-0006-8677-7495}}
\affiliation{INFN, Sezione di Genova, 16147 Genova, Italia}
\affiliation{Universit\`a degli Studi di Genova, 16126 Genova, Italia}
\author{N.~Charitonidis\orcidlink{0000-0001-9506-1022}}
\affiliation{CERN, European Organization for Nuclear Research, CH-1211 Geneva, Switzerland}
\author{P.~Crivelli\orcidlink{0000-0001-5430-9394}}
\email[\textbf{e-mail}:]{paolo.crivelli@cern.ch}
\affiliation{ETH Z\"urich, Institute for Particle Physics and Astrophysics, CH-8093 Z\"urich, Switzerland}
\author{E.~Depero\orcidlink{0000-0003-2239-1746}}
\affiliation{ETH Z\"urich, Institute for Particle Physics and Astrophysics, CH-8093 Z\"urich, Switzerland}
\author{A.~V.~Dermenev\orcidlink{0000-0001-5619-376X}}
\affiliation{Authors affiliated with an institute covered by a cooperation agreement with CERN}
\author{S.~V.~Donskov\orcidlink{0000-0002-3988-7687}}
\affiliation{Authors affiliated with an institute covered by a cooperation agreement with CERN}
\author{R.~R.~Dusaev\orcidlink{0000-0002-6147-8038}}
\affiliation{Authors affiliated with an institute covered by a cooperation agreement with CERN}
\author{T.~Enik\orcidlink{0000-0002-2761-9730}}
\affiliation{Authors affiliated with an international laboratory covered by a cooperation agreement with CERN}
\author{V.~N.~Frolov}
\affiliation{Authors affiliated with an international laboratory covered by a cooperation agreement with CERN}
\author{R.~B.~Galleguillos~Silva}
\affiliation{Center for Theoretical and Experimental Particle Physics, Facultad de Ciencias Exactas, Universidad Andres Bello, Fernandez Concha 700, Santiago, Chile}
\affiliation{Millennium Institute for Subatomic Physics at High-Energy Frontier (SAPHIR), Fernandez Concha 700, Santiago, Chile}
\author{A.~Gardikiotis\orcidlink{0000-0002-4435-2695}}
\affiliation{Physics Department, University of Patras, 265 04 Patras, Greece}
\author{S.~V.~Gertsenberger\orcidlink{0009-0006-1640-9443}}
\affiliation{Authors affiliated with an international laboratory covered by a cooperation agreement with CERN}
\author{S. Girod}
\affiliation{CERN, European Organization for Nuclear Research, CH-1211 Geneva, Switzerland}
\author{S.~N.~Gninenko\orcidlink{0000-0001-6495-7619}}
\affiliation{Authors affiliated with an institute covered by a cooperation agreement with CERN}
\affiliation{Center for Theoretical and Experimental Particle Physics, Facultad de Ciencias Exactas, Universidad Andres Bello, Fernandez Concha 700, Santiago, Chile}
\author{M.~H\"osgen}
\affiliation{Universit\"at Bonn, Helmholtz-Institut f\"ur Strahlen-und Kernphysik, 53115 Bonn, Germany}
\author{V.~A.~Kachanov\orcidlink{0000-0002-3062-010X}}
\affiliation{Authors affiliated with an institute covered by a cooperation agreement with CERN}
\author{Y.~Kambar\orcidlink{0009-0000-9185-2353}}
\affiliation{Authors affiliated with an international laboratory covered by a cooperation agreement with CERN}
\author{A.~E.~Karneyeu\orcidlink{0000-0001-9983-1004}}
\affiliation{Authors affiliated with an institute covered by a cooperation agreement with CERN}
\author{E.~A.~Kasianova}
\affiliation{Authors affiliated with an international laboratory covered by a cooperation agreement with CERN}
\author{G.~Kekelidze\orcidlink{0000-0002-5393-9199}}
\affiliation{Authors affiliated with an international laboratory covered by a cooperation agreement with CERN}
\author{B.~Ketzer\orcidlink{0000-0002-3493-3891}}
\affiliation{Universit\"at Bonn, Helmholtz-Institut f\"ur Strahlen-und Kernphysik, 53115 Bonn, Germany}
\author{D.~V.~Kirpichnikov\orcidlink{0000-0002-7177-077X}}
\affiliation{Authors affiliated with an institute covered by a cooperation agreement with CERN}
\author{M.~M.~Kirsanov\orcidlink{0000-0002-8879-6538}}
\affiliation{Authors affiliated with an institute covered by a cooperation agreement with CERN}
\author{V.~N.~Kolosov}
\affiliation{Authors affiliated with an institute covered by a cooperation agreement with CERN}
\author{V.~A.~Kramarenko\orcidlink{0000-0002-8625-5586}}
\affiliation{Authors affiliated with an institute covered by a cooperation agreement with CERN}
\affiliation{Authors affiliated with an international laboratory covered by a cooperation agreement with CERN}
\author{L.~V.~Kravchuk\orcidlink{0000-0001-8631-4200}}
\affiliation{Authors affiliated with an institute covered by a cooperation agreement with CERN}
\author{N.~V.~Krasnikov\orcidlink{0000-0002-8717-6492}}
\affiliation{Authors affiliated with an institute covered by a cooperation agreement with CERN}
\affiliation{Authors affiliated with an international laboratory covered by a cooperation agreement with CERN}
\author{S.~V.~Kuleshov\orcidlink{0000-0002-3065-326X}}
\affiliation{Center for Theoretical and Experimental Particle Physics, Facultad de Ciencias Exactas, Universidad Andres Bello, Fernandez Concha 700, Santiago, Chile}
\affiliation{Millennium Institute for Subatomic Physics at High-Energy Frontier (SAPHIR), Fernandez Concha 700, Santiago, Chile}
\author{V.~E.~Lyubovitskij\orcidlink{0000-0001-7467-572X}}
\affiliation{Authors affiliated with an institute covered by a cooperation agreement with CERN}
\affiliation{Universidad T\'ecnica Federico Santa Mar\'ia and CCTVal, 2390123 Valpara\'iso, Chile}
\affiliation{Millennium Institute for Subatomic Physics at High-Energy Frontier (SAPHIR), Fernandez Concha 700, Santiago, Chile}
\author{V.~Lysan\orcidlink{0009-0004-1795-1651}}
\affiliation{Authors affiliated with an international laboratory covered by a cooperation agreement with CERN}
\author{V.~A.~Matveev\orcidlink{0000-0002-2745-5908}}
\affiliation{Authors affiliated with an international laboratory covered by a cooperation agreement with CERN}
\author{R.~Mena~Fredes}
\affiliation{Millennium Institute for Subatomic Physics at High-Energy Frontier (SAPHIR), Fernandez Concha 700, Santiago, Chile}
\affiliation{Universidad T\'ecnica Federico Santa Mar\'ia and CCTVal, 2390123 Valpara\'iso, Chile}
\author{R.~ G.~Mena~Yanssen}
\affiliation{Millennium Institute for Subatomic Physics at High-Energy Frontier (SAPHIR), Fernandez Concha 700, Santiago, Chile}
\affiliation{Universidad T\'ecnica Federico Santa Mar\'ia and CCTVal, 2390123 Valpara\'iso, Chile}
\author{L.~Molina Bueno\orcidlink{0000-0001-9720-9764}}
\email[\textbf{e-mail}:]{laura.molina.bueno@cern.ch}
\affiliation{Instituto de Fisica Corpuscular (CSIC/UV), Carrer del Catedratic Jose Beltran Martinez, 2, 46980 Paterna, Valencia, Spain}
\author{M.~Mongillo\orcidlink{0009-0000-7331-4076}}
\affiliation{ETH Z\"urich, Institute for Particle Physics and Astrophysics, CH-8093 Z\"urich, Switzerland}
\author{D.~V.~Peshekhonov\orcidlink{0009-0008-9018-5884}}
\affiliation{Authors affiliated with an international laboratory covered by a cooperation agreement with CERN}
\author{V.~A.~Polyakov\orcidlink{0000-0001-5989-0990}}
\affiliation{Authors affiliated with an institute covered by a cooperation agreement with CERN}
\author{B.~Radics\orcidlink{0000-0002-8978-1725}}
\affiliation{York University, Toronto, Canada}
\author{K.~M.~Salamatin\orcidlink{0000-0001-6287-8685}}
\affiliation{Authors affiliated with an international laboratory covered by a cooperation agreement with CERN}
\author{V.~D.~Samoylenko}
\affiliation{Authors affiliated with an institute covered by a cooperation agreement with CERN}
\author{D.~A.~Shchukin\orcidlink{0009-0007-5508-3615}}
\affiliation{Authors affiliated with an institute covered by a cooperation agreement with CERN}
\author{O.~Soto}
\affiliation{Departamento de Fisica, Facultad de Ciencias, Universidad de La Serena, Avenida Cisternas 1200, La Serena, Chile}
\affiliation{Millennium Institute for Subatomic Physics at High-Energy Frontier (SAPHIR), Fernandez Concha 700, Santiago, Chile}
\author{H.~Sieber\orcidlink{0000-0003-1476-4258}}
\email[\textbf{e-mail}:]{henri.hugo.sieber@cern.ch}
\affiliation{ETH Z\"urich, Institute for Particle Physics and Astrophysics, CH-8093 Z\"urich, Switzerland}
\author{V.~O.~Tikhomirov\orcidlink{0000-0002-9634-0581}}
\affiliation{Authors affiliated with an institute covered by a cooperation agreement with CERN}
\author{I.~V.~Tlisova\orcidlink{0000-0003-1552-2015}}
\affiliation{Authors affiliated with an institute covered by a cooperation agreement with CERN}
\author{A.~N.~Toropin\orcidlink{0000-0002-2106-4041}}
\affiliation{Authors affiliated with an institute covered by a cooperation agreement with CERN}
\author{M.~Tuzi\orcidlink{0009-0000-6276-1401}}
\affiliation{Instituto de Fisica Corpuscular (CSIC/UV), Carrer del Catedratic Jose Beltran Martinez, 2, 46980 Paterna, Valencia, Spain}
\author{M. B. Veit}
\affiliation{Johannes Gutenberg Universitaet Mainz, Germany}
\author{P.~V.~Volkov\orcidlink{0000-0002-7668-3691}}
\affiliation{Authors affiliated with an international laboratory covered by a cooperation agreement with CERN}
\author{V.~Yu.~Volkov\orcidlink{0009-0005-3500-5121}}
\affiliation{Authors affiliated with an institute covered by a cooperation agreement with CERN}
\author{I.~V.~Voronchikhin\orcidlink{0000-0003-3037-636X}}
\affiliation{Authors affiliated with an institute covered by a cooperation agreement with CERN}
\author{J.~Zamora-Sa\'a\orcidlink{0000-0002-5030-7516}}
\affiliation{Center for Theoretical and Experimental Particle Physics, Facultad de Ciencias Exactas, Universidad Andres Bello, Fernandez Concha 700, Santiago, Chile}
\affiliation{Millennium Institute for Subatomic Physics at High-Energy Frontier (SAPHIR), Fernandez Concha 700, Santiago, Chile}
\author{A.~S.~Zhevlakov\orcidlink{0000-0002-7775-5917}}
\affiliation{Authors affiliated with an international laboratory covered by a cooperation agreement with CERN}

\begin{abstract}
We report on a search for a new $Z'$ ($L_\mu-L_\tau$) vector boson performed at the NA64 experiment employing a high energy muon beam and a missing energy-momentum technique. 
Muons from the M2 beamline at the CERN Super Proton Synchrotron with a momentum of 160 GeV/c are directed to an active target. A signal event is a single scattered muon with momentum $<$ 80 GeV/c in the final state, accompanied by missing energy, i.e.  no detectable activity in the downstream calorimeters. 
For a total statistic of $(1.98\pm0.02)\times10^{10}$ muons on target, no event is observed in the expected signal region. This allows us to set new limits on part of the remaining $(m_{Z'},\ g_{Z'})$ parameter space which could provide an explanation for the muon $(g-2)_\mu$ anomaly. Additionally, our study excludes part of the parameter space suggested by the thermal Dark Matter relic abundance.
Our results pave the way to explore Dark Sectors and light Dark Matter with muon beams in a unique and complementary way to other experiments. 

\end{abstract}
\maketitle
\begin{figure}[H]
    \centering
    \includegraphics[width=0.38\textwidth]{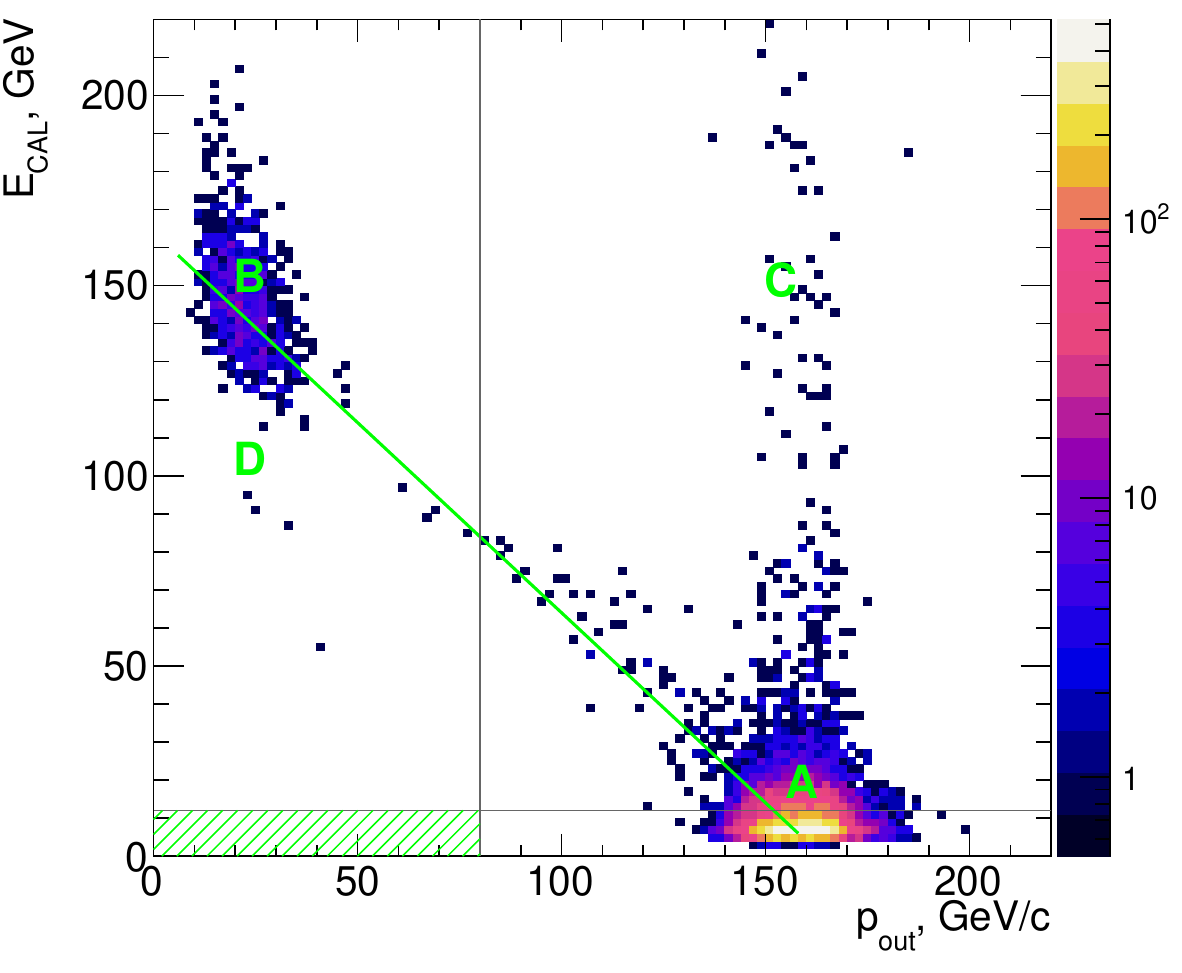}
    \caption{\label{fig:bi-plot}Event distribution in the $(p_\text{out},\ E_\text{CAL})$ plane before the MIP-compatible requirement selection criterion. The signal box is defined as the shaded green rectangular area and the controlled region labelled with $A$ through $D$ (see text).}
\end{figure}
\begin{figure}[H]
    \centering
    \includegraphics[width=0.45\textwidth]{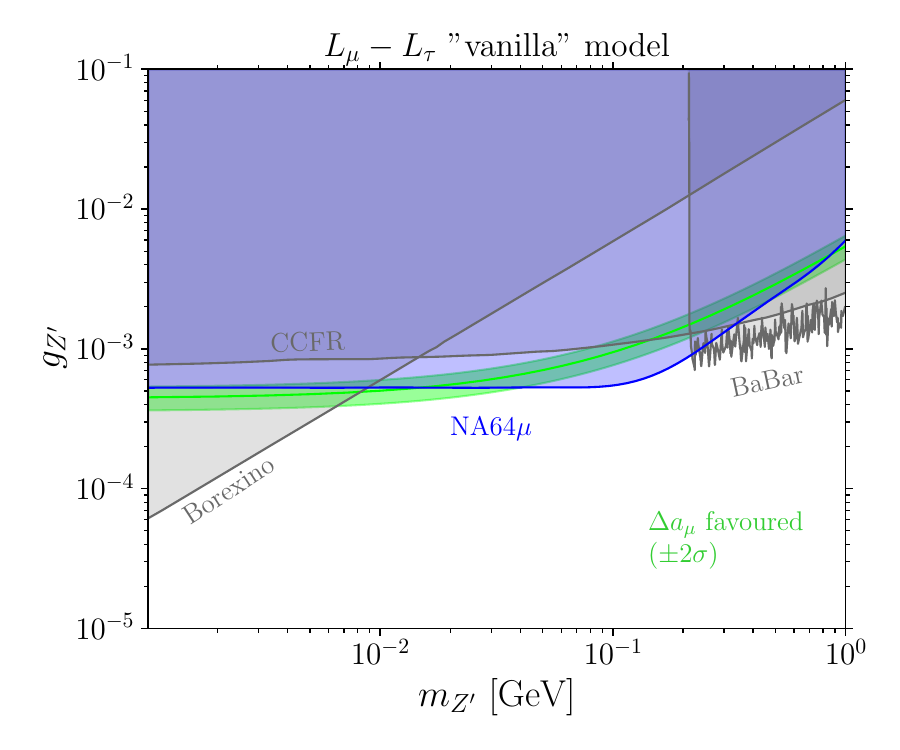}
    \caption{NA64$\mu$ 90\% CL exclusion limits on the coupling $g_{Z^\prime}$ as a function of the $Z'$ mass, $m_{Z'}$, for the vanilla $L_\mu-L_\tau$ model. The $\pm2\sigma$ band for the $Z'$ contribution to the $(g-2)_\mu$ discrepancy is also shown. Existing constraints from BaBar \cite{Godang:2016gna,Capdevilla:2021kcf} and from neutrino experiments such as BOREXINO \cite{Kamada:2015era,Kaneta:2016uyt,Gninenko:2020xys} and CCFR \cite{Altmannshofer:2014pba,CCFR:1991lpl} are plotted.}
   \label{fig:g-2-vanilla}
\end{figure}
\begin{figure}[H]
    \centering
    \includegraphics[width=0.45\textwidth]{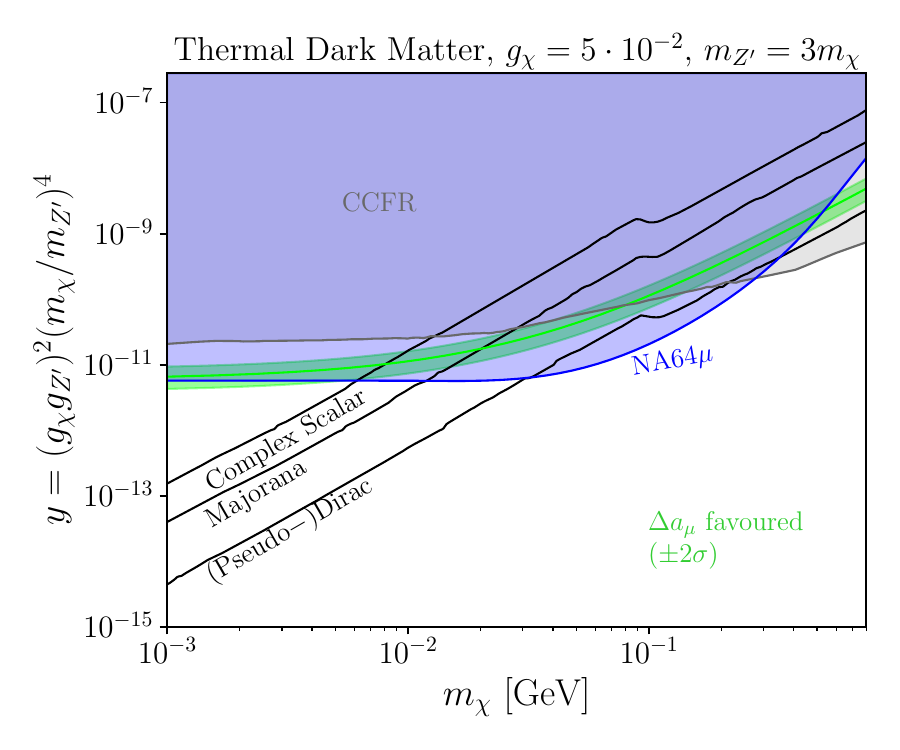}
    \caption{The 90\% CL exclusion limits obtained by the NA64$\mu$ experiment in the $(m_\chi,\ y)$ parameters space for thermal Dark Matter charged under $U(1)_{L_\mu-L_\tau}$ with $m_{Z'}=3m_\chi$ and the coupling  $g_\chi=5\times10^{-2}$ for $2\times10^{10}$ MOT.  The branching ratio to invisible final states is assumed to be $\text{Br}(Z'\rightarrow\text{invisible})\simeq1$ (see text for details). Existing bounds obtained through the CCFR experiment \cite{CCFR:1991lpl,Altmannshofer:2014pba} are shown for completeness. The thermal targets for the different scenarios are taken from \cite{Berlin:2018bsc}.}
    \label{fig:thermal-dm-limits}
\end{figure}

Dark Sectors (DS) are a promising paradigm to address open questions of the Standard Model (SM) such as the Dark Matter (DM) origin~\cite{Feng:2010gw}. In this framework, one postulates a new sector of particles below the electroweak scale that are not charged under the SM but could have a phenomenology of their own \cite{ArkaniHamed:2008qn,Pospelov:2008jd,Hooper:2012cw,Pospelov:2007mp,Pospelov:2008zw,Essig:2013lka}.
In addition to gravity, the interactions between DS states and the SM could proceed through portal mediators~\cite{Kobzarev:1966qya,Blinnikov:1982eh,Foot:1991bp,Hodges:1993yb,Berezhiani:1995am}.
If one assumes that DM is made by the lightest stable DS particles, the resulting feeble interaction between the two sectors is compatible with cosmological observations and, thus, would accommodate a solution to the DM problem. DS models became an extremely fertile domain of exploration with many different techniques tackling the very large parameter space of possible DM candidates (see e.g. for recent reviews \cite{Jaeckel:2020dxj,Lanfranchi:2020crw,Krnjaic:2022ozp,Antel:2023hkf}).
Models with lepton numbers $L_\mu-L_\tau$ gauging are very attractive to explain the origin of DM and, at the same time, provide an explanation for the long-standing $g-2$ muon anomaly \cite{Holst:2021lzm}. 
The $Z'$ vector boson originates from the broken $U(1)_{L_\mu-L_\tau}$ symmetry and couples directly to the second and third lepton generations, and their corresponding left-handed neutrinos through the coupling $g_{Z'}$ \cite{He:1990pn,He:1991qd,Foot:1994vd,Altmannshofer:2016jzy,Kile:2014jea,Park:2015gdo}. 
The extension of this model to interactions with DM candidates, being consistent in predicting
the observed DM relic density \cite{Feng:2008mu,Feng:2008ya,Arcadi:2017kky,Planck:2018vyg} 
, is achieved by adding to the Lagrangian a term of the type $\mathcal{L}\supseteq -g_\chi Z_\mu^\prime J_\chi^\mu$ with the current $J_\chi^\mu$  and the coupling $g_\chi$ of the $Z_\mu^\prime$ to the DM candidates. In the case where $m_{Z^\prime}>m_\chi$ (away from the near on-shell resonant enhancement $m_{Z^\prime}\simeq2m_{\chi}$), the relic density is driven by $\bar{\chi}\chi(\rightarrow Z^{(\ast)\prime}\rightarrow)\bar{f}f$, $f=\mu,\tau,\nu$, with the relevant $s-$channel annihilation cross-section scaling as \cite{Altmannshofer:2016jzy,Arcadi:2017kky} $\langle\sigma v\rangle\propto(g_\chi g_{Z'})^2m_\chi^2/m_{Z^\prime}^4=y m_\chi^{-2}$. Below the resonance, $m_{Z'}<2m_\chi$, the $t-$channel annihilation is $\bar{\chi}\chi\rightarrow Z^\prime Z^\prime$, with $\langle\sigma v\rangle\propto g_{\chi}^4/m_\chi^2$.\\ \indent
Within this framework, the discrepancy between the experimental \cite{Muong-2:2023cdq} and SM predicted \cite{Davier:2017zfy,Keshavarzi:2018mgv,Colangelo:2018mtw,Hoferichter:2019mqg,Davier:2019can,Keshavarzi:2019abf,Kurz:2014wya,Melnikov:2003xd,Masjuan:2017tvw,Colangelo:2017fiz,Hoferichter:2018kwz,Gerardin:2019vio,Bijnens:2019ghy,Colangelo:2019uex,Colangelo:2014qya,Blum:2019ugy,Aoyama:2012wk,Aoyama:2019ryr,Gnendiger:2013pva} $(g-2)_\mu$ values can also be explained through loop corrections \cite{Pospelov:2007mp,Gninenko:2014pea,Gninenko:2001hx,Chen:2017awl,Gninenko:2018tlp,Kirpichnikov:2020tcf,Amaral:2021rzw}.
The current bounds for $m_{Z^\prime}>2m_\mu$ arise from direct searches, sensitive to the kinematically allowed visible decay channel $Z^\prime\rightarrow\mu^{+}\mu^{-}$ \cite{BaBar:2016sci,CMS:2018yxg,ATLAS:2023vxg,Kamada:2015era}. Neutrino scattering experiments
\cite{CCFR:1991lpl,CHARM-II:1990dvf} and missing energy searches through $Z'\rightarrow\bar{\chi}\chi$ \cite{NA64:2022rme,Belle-II:2022yaw} provide constraints for $m_{Z^\prime}<2m_\mu$. The lower bound is set through the $Z'$ contribution to the radiation density of the Universe through $\Delta N_\text{eff}$, with its value being defined from both the CMB spectrum \cite{Planck:2018vyg} and the Big Bang nucleosynthesis (BBN) \cite{Ahlgren:2013wba,Kamada:2015era,Escudero:2019gzq} to $m_{Z'}>3-10$ MeV \cite{Sabti:2019mhn} and $g_{Z^\prime}\sim10^{-4}-10^{-3}$. \\ \indent 
In this Letter, we report on the first results of the NA64 experiment muon program, dubbed NA64$\mu$, looking for Dark Sectors weakly coupled to muons. 
The experimental set-up and working principle are schematically shown in Fig. \ref{fig:na64mu-setup}.
\begin{widetext}
\begin{minipage}{\linewidth} 
\begin{figure}[H]
    \centering
    \includegraphics[width=0.99\textwidth]{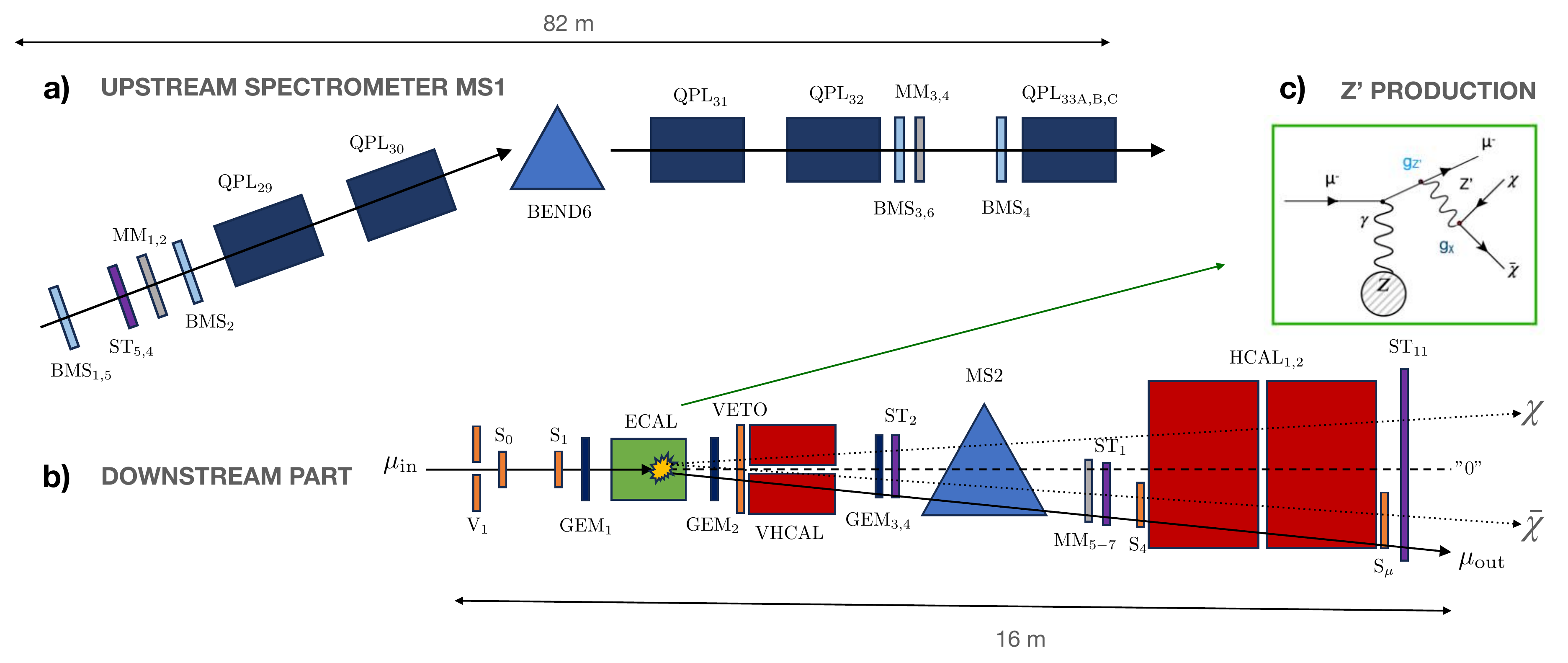}
    \caption{\label{fig:na64mu-setup}Schematic illustration of the NA64$\mu$ set-up. a) The spectrometer in the \emph{upstream} region is used for identifying incoming muons with momentum $p_\text{in}\simeq160$ GeV/c. b) The \emph{downstream} part composed of calorimeters and a second spectrometer measures the momentum of the scattered muons to search for the $Z'$ vector boson production. c) Sketch of the bremsstrahlung-like reaction $\mu N\rightarrow\mu N(Z^\prime\rightarrow\text{invisible})$ of 160 GeV/c incident muons on the ECAL target.}
\end{figure}
\end{minipage}
\end{widetext}
If a $Z'$ boson exists, it could be produced in the bremsstrahlung-like reaction of a high energy muon scattering off atomic nuclei in a target ($N$), followed by its prompt invisible decay $\mu N\rightarrow\mu N Z';Z'\rightarrow\text{invisible}$, with only $Z'\rightarrow\bar{\nu}\nu$ in the \emph{vanilla} model, and additionally, $Z'\rightarrow\bar{\chi}\chi$ for DM candidates \cite{Gninenko:2017yus,Kirpichnikov:2021jev,Sieber:2023nkq}.  For a value of $g_\chi = 5\times 10^{-2}$ one can accommodate in the same parameter space the muon g-2 and the DM relic prediction \cite{Kahn:2018cqs}. In the region of interest (below $m_\chi<$1 GeV) $g_\chi\gg g_{Z'}$, the branching ratio to DS invisible final states can be assumed to be $\text{Br}(Z'\rightarrow\bar{\chi}\chi)\simeq1$, while the ones in visible states ($Z'\rightarrow \mu^+\mu^-$) and neutrinos can be neglected.
\\ \indent
The search for signal events is based on a missing energy-momentum technique which consists of the detection of a primary beam muon with a momentum of 160 GeV/c in the initial state, and a single muon scattered off an active target with missing momentum $>$ 80 GeV/c in the final state, accompanied by missing energy,  i.e. no detectable electromagnetic or hadronic activity in the downstream calorimeters.\\ \indent
 The 160 GeV/c muons are delivered by the M2 beamline at the CERN Super Proton Synchrotron (SPS)\cite{Doble:1994np}. The beam optics comprises a series of quadrupoles focusing the beam before the target with a divergency $\sigma_x\sim 0.9$ and $\sigma_y\sim 1.9$ cm \cite{Sieber:2021fue}. The incoming muon momentum is reconstructed through a magnetic spectrometer (MS1) consisting of three 5 T$\cdot$m bending magnets (BEND6), together with four micro-mesh gas detectors (Micromegas, MM$_{1-4}$), two straw tubes chambers (ST$_{5,4}$) and six scintillator hodoscopes, the beam momentum stations (BMS$_{1-6}$). The obtained resolution is $\sigma_{p_\text{in}} /p_\text{in}\simeq3.8\%$. 
The target is an active electromagnetic calorimeter (ECAL) 
composed of Shashlik-type modules made of a lead-scintillator (Pb-Sc) resulting in 40 radiation lengths ($X_0$). 
The ECAL is followed by a large $55\times55$ cm$^2$ high-efficiency veto counter (VETO) and a 5 nuclear interaction lengths ($\lambda_\text{int}$) copper-Sc (Cu-Sc) hadronic calorimeter (VHCAL) with a hole in its middle. The outgoing muon momentum is reconstructed through a second magnetic spectrometer consisting of a single 1.4 T$\cdot$m bending magnet (MS2) together with four gaseous electron multiplier trackers (GEM$_{1-4}$), two additional straw chambers (ST$_{2,1}$) and three 20$\times$8 cm$^2$ Micromegas (MM$_{5-7}$) yielding a resolution of $\sigma_{p_{\text{out}}}/p_\text{out}\simeq4.4\%$. To identify and remove any residuals from interactions in the detectors upstream MS2 and ensure maximal hermeticity, two large 120$\times$60 cm$^2$, $\lambda_\text{int}\simeq\ 30$ iron-Sc (Fe-Sc) HCAL modules (HCAL$_{1,2}$) are placed at the end of the set-up together with a 120$\times$60 cm$^2$ UV straw, ST$_{11}$. The trigger system is defined by a veto counter with a hole (V$_1$) and a set of Sc counters (S$_{0-1}$) before the target, together with two $20\times20$ cm$^2$ and $30\times30$ cm$^2$ Sc counters (S$_{4}$ and S$_{\mu}$) sandwiching the HCAL modules, shifted from the undeflected beam axis (referred to as \emph{zero-line}) to detect the scattered muons.\\ \indent
The data were collected in two trigger configurations (S$_0\times$S$_1\times\overline{\text{V}_1}\times$S$_4\times$S$_\mu$) with different S$_4$ and S$_\mu$ distances to the zero-line along the deflection axis $\hat{x}$, namely S$_\mu\hat{x}=-152$ mm and S$_\mu\hat{x}=-117$ mm with a similar S$_4\hat{x}=-65$ mm. The corresponding measured rate is $0.04\%$ and $0.07\%$ of the calibration trigger (S$_{0,1}\times\overline{\text{V}_1}$) coincidences at a beam intensity of $2.8\times10^{6}$ $\mu/$spill. In each configuration, we recorded respectively $(11.7\pm0.1)\times10^{9}$ and $(8.1\pm0.1)\times10^{9}$ muons on target (MOT) yielding a total accumulated statistics of $(1.98\pm0.02)\times10^{10}$ MOT.\\ \indent
A detailed \texttt{GEANT4}-based \cite{GEANT4:2002zbu,Allison:2016lfl} Monte Carlo (MC) simulation is performed to study the main background sources and the response of the detectors and the muon propagation. In the latter case, the full beam optics developed by the CERN BE-EA beam department is encompassed in the simulation framework using separately both the \texttt{TRANSPORT}, \texttt{HALO} and \texttt{TURTLE} programs \cite{Brown:1983jnh,Iselin:1974fu,Brown:1974ns}, as well the \texttt{GEANT4} compatible beam delivery simulation (\texttt{BDSIM}) program \cite{Nevay:2018zhp,Nevay:2018xwd,Nevay:2019kmu} to simulate secondaries interactions in the beamline material. The signal acceptance is carefully studied using the \texttt{GEANT4} interface \texttt{DMG4} package \cite{Bondi:2021nfp}, including light mediators production cross-sections computations through muon bremsstrahlung \cite{Sieber:2023nkq}. The placements of $\text{S}_4$ and $\text{S}_\mu$ are optimized to compensate for the low signal yield at high masses, $\sigma_{Z^\prime}\sim g_{Z^\prime}^2\alpha Z^2/m_{Z^\prime}^2$, with $\alpha$ the fine structure constant and $Z$ the atomic number of the target, through angular acceptance being maximized for a scattered muon angle $\psi_\mu^\prime\sim10^{-2}$ rad after ECAL. In addition, the trigger counters downstream of MS2 account for the expected 160 GeV/c mean deflected position at the level of S$_4$, estimated at $\langle\delta x\rangle\simeq-12.0$ mm from a detailed \texttt{GenFit}-based \cite{Rauch:2014wta,Bilka:2019ang} Runge-Kutta (RK) extrapolation scheme.\\ \indent
The signal box, $p_\text{out}^\text{cut}\leq80$ GeV/c and $E_\text{CAL}^\text{cut}<12$ GeV, is optimized with signal simulations and data to maximize the sensitivity. The cut on the total energy deposit in the calorimeters, $E_\text{CAL}^\text{cut}$, is obtained from the sum of the minimum ionizing particle (MIP) peaks of the related energy spectra. \\ \indent 
To minimize the background, the following set of selection criteria is used. (i) The incoming momentum should be in the momentum range $160\pm20$ GeV/c. (ii) A single track is reconstructed in each magnetic spectrometer (MS1 and MS2) to ensure that a single muon traverses the full set-up. (iii) At most one hit is reconstructed in MM$_{5-7}$ and ST$_1$ (no multiple hits) and the corresponding extrapolated track to the HCAL face is compatible with a MIP energy deposit in the expected cell. This cut verifies that no energetic enough secondaries from interactions upstream MS2 arrive at the HCAL. (vi) The energy deposit in the calorimeters and the veto should be compatible with a MIP. This cut enforces the selection of events with no muon nuclear interactions in the calorimeters. The aforementioned cut-flow is applied to events distributed in the outgoing muon momentum and total energy deposit plane, $(p_\text{out},\ E_\text{CAL})$, as shown in Fig. \ref{fig:bi-plot}. \\ \indent
Region $A$ is inherent to events with MIP-compatible energy deposits in all of the calorimeters, resulting in $p_\text{in}\simeq p_\text{out}\simeq160$ GeV/c. By design, most of unscattered beam muons do not pass through the S$_{4}$ and S$_\mu$ counters, however, the trigger condition can be fulfilled by sufficiently energetic residual ionization $\mu N\rightarrow\mu N+\delta e$ originating from the downstream trackers MM$_{5-7}$ or last HCAL$_{2}$ layers. The accumulation of events in region $C$ is associated with large energy deposition of the full-momentum scattered muon in the HCAL, while region $B$ corresponds to a hard scattering/bremsstrahlung in the ECAL, with a soft outgoing muon and full energy deposition in either the active target or HCAL. The small number of events between $p_\text{out}\geq50$ GeV/c and $p_\text{out}\leq100$ GeV/c associated with hard muon bremsstrahlung events, $\mu N\rightarrow\mu N+\gamma$, with $\psi_\mu^\prime\ll10^{-2}$ rad, is a result of the trigger optimization for signal events emitted at larger angles. 
The events in the region $D$ are associated with muon nuclear interactions in the ECAL, $\mu N\rightarrow\mu+X$, with $X$ containing any combination of $\pi'$s, $K$, $p$, $n$..., with low-energy charged hadrons being deflected away in MS2, going out of the detector acceptance (typically the HCAL modules).\\ \indent
\begin{table}[H]
    \centering
    \resizebox{0.42\textwidth}{!}{
    \begin{tabular}{lr}
    \hline
    \hline
    Background source & Background, $n_b$ \\
    \hline
    (I) Momentum mis-reconstruction & $0.05\pm0.03$ \\
    (II) $K\rightarrow\mu+\nu$, ... in-flight decays & $0.010\pm0.001$ \\
    (III) Calorimeter non-hermeticity & $<0.01$ \\
    \hline
    Total $n_b$ (conservatively) & $0.07\pm0.03$ \\
    \hline
    \hline
    \end{tabular}
    }
    \caption{\label{tab:background}Expected main background level within the signal box, together with its statistical error, for the 2022 muon pilot run corresponding to $\sim2\times10^{10}$ MOT.}
\end{table}
An exhaustive discussion of background sources is given in \cite{Gninenko:2640930,Sieber:2021fue}. The main processes are summarised in Table \ref{tab:background}, with the dominant background contribution being associated with (I) momentum mis-reconstruction of the scattered muon in MS2. An incoming muon with 160 GeV/c is reconstructed after the target with momentum $\leq80$ GeV/c, whereas it truly is 160 GeV/c. This background is evaluated from data by selecting a sample of incoming muons within a $\sim2\sigma_{p_\text{in}}$ window around its nominal momentum $\langle p_\text{in}\rangle=160$ GeV/c and extrapolating the tails of the corresponding downstream momentum distribution $p_\text{out}$ towards 80 GeV/c. The second most important background process is (II) kaons decays to (semi-)leptonic final states with muons, $K\rightarrow\mu\nu,\ ...$, before the ECAL target. Because of the level of hadron contamination in the M2 beamline, $P_h\simeq5\times10^{-5}$ \cite{Doble:1994np}, incoming kaons could be reconstructed through MS1 with a momentum passing the selection criterion (i) and subsequently decaying to muons with energy $\leq$ 80 GeV, with the neutrino carrying away the remaining energy. This contribution is estimated from MC with the hadron contamination being extracted from existing data \cite{Doble:1994np}. Pion decays do not contribute to this background since due to kinematics the muon momentum is always $\geq$ 80 GeV.  Another background source is associated with (III) non-hermeticity in the calorimeters due to muon nuclear interactions in the target. As such, a leading hadron with energy $E_h\geq80$ GeV could be produced and escape the ECAL with lesser energetic charged secondaries and the scattered muon. Because of the non-zero charge of the particles and the trigger acceptance, low-energy secondaries are deflected away through MS2 resulting in missing energy events. This background is extrapolated to the signal region from region D of Fig. \ref{fig:bi-plot}. After applying all selection criteria (i-iv) and summing up the processes contributing to the background, the expected background level is found to be $0.07\pm0.03$ for the total statistics of $\sim2\times10^{10}$ MOT.\\ \indent
The upper limits on the coupling $g_{Z'}$ as a function of its mass $m_{Z'}$ are estimated at 90\% confidence level (CL) following the modified frequentist approach. In particular, the \texttt{RooFit}/\texttt{RooStats}-based \cite{Verkerke:2003ir,Wolffs:2022fkh,Moneta:2010pm} profile likelihood ratio statistical test is used in the asymptotic approximation \cite{Cowan:2010js}. The total number of signal events falling within the signal box is given by the sum of the two trigger configurations $t$
\begin{equation}
N_{Z'}=\sum_{t=1,2}N_{Z'}^{t}=\sum_{t=1,2}N_\text{MOT}^{t}\times\epsilon_{Z'}^{t}\times N_{Z'}^{t}(m_{Z'},g_{Z'}),
\label{eq:signal-yield}
\end{equation}
where $N_\text{MOT}^{t}$ is the number of MOT for trigger configuration $t$,  $N_{Z}^{t}$ the number of signals per MOT produced in the ECAL target, depending on the mass/coupling parameters $m_{Z'}$ and $g_{Z'}$, and $\epsilon_{Z'}^{t}$ the trigger-dependent signal efficiency. \\ \indent
The main systematic effects contributing to the signal yield defined in Eq. \eqref{eq:signal-yield} are studied in detail. The uncertainty on $N_\text{MOT}^{t}$ is conservatively set to 1\%. The systematics associated with the $Z'$ production cross-section are extracted from the uncertainty introduced by the Weisz\"{a}cker-Williams (WW) approximation and from QED corrections to the exact tree-level (ETL) expression. In the former case, the relative error in assessing the number of produced $Z'$ ($N_{Z'}^{t}$) is found to be $2\%$ \cite{Kirpichnikov:2021jev,Sieber:2023nkq}. In the latter case, both the running of $\alpha$ at the upper bound $Q^2\simeq m_{Z^\prime}\sim\mathcal{O}(1)$ GeV and higher order corrections from soft photon emissions are estimated to contribute through respectively $\Delta N_{Z'}\sim\alpha^2g_{Z'}^2Z^2$ and through the Sudakov factor $\Delta N_\text{soft}\sim\exp(-\alpha/\pi)$ at the level of 2.4\% and 1.4\%. Uncertainties relative to the Pb purity of the ECAL target are addressed at the level of 1\%. The systematics on $\epsilon_{Z'}^{t}$ are evaluated by comparing the detector responses in MC and data around the MIP-compatible peak, in particular in the ECAL and HCAL. Through spectra integration and peak ratio, it is found that the related cumulative uncertainty does not exceed 4\%. Because of the strong dependence of the efficiency $\epsilon_{Z'}^{t}$ on the trigger configuration $t$, in particular on the distance from the zero-line, additional uncertainties due to S$_4$ and S$_\mu$ misalignment are studied through the change in efficiency as a response to small displacements of the Sc counters. Because of the $m_{Z'}$ mass-dependence of the trigger rate \cite{Sieber:2021fue}, the resulting uncertainty reaches up to $\leq5\%$. As such the total systematic in the signal yield of Eq. \eqref{eq:signal-yield} is $\leq8\%$ 
. The acceptance loss due to accidentals (pile-up events, $\sim13\%$) entering the trigger time window is taken into account in the final efficiency computations. The signal efficiency peaks at its maximum of $\sim12\%$ for the mass range $\mathcal{O}(100\ \text{MeV}-1\ \text{GeV})$.\\ \indent
After unblinding, no event compatible with $Z'$ production is found in the signal box. This allows us to set the 90\% CL exclusion limits on $g_{Z^\prime}$ which are plotted in Fig. \ref{fig:g-2-vanilla} in the $(m_{Z^\prime},\ g_{Z^\prime})$ parameter space, together with the values of $\Delta a_\mu$ compatible with the muon $g-2$ anomaly, within $\pm2\sigma$. The band is computed using the latest results of the Muon $g-2$ collaboration for the combined Runs 2 and 3 (2019-2020), $a_\mu(\text{Exp})=116\ 592\ 059(22)\times10^{-11}$ \cite{Muong-2:2023cdq} and the SM prediction of  $a_\mu(\text{SM})=116\ 591\ 810(43)\times10^{-11}$  from the Muon $g-2$ Theory Initiative (TI) \cite{Aoyama:2020ynm,Davier:2017zfy,Keshavarzi:2018mgv,Colangelo:2018mtw,Hoferichter:2019mqg,Davier:2019can,Keshavarzi:2019abf,Kurz:2014wya,Melnikov:2003xd,Masjuan:2017tvw,Colangelo:2017fiz,Hoferichter:2018kwz,Gerardin:2019vio,Bijnens:2019ghy,Colangelo:2019uex,Colangelo:2014qya,Blum:2019ugy,Aoyama:2012wk,Aoyama:2019ryr,Gnendiger:2013pva}.
It is worth noticing that the latest results from the CMD-3 collaboration \cite{CMD-3:2023alj,CMD-3:2023rfe} on the $\pi^{+}\pi^{-}$ disagree within the 2.5-5$\sigma$ level with the TI value and recent lattice QCD computations from the BMW collaboration \cite{Borsanyi:2020mff} are in tension by 2.1$\sigma$. 
Our results, excluding masses $m_{Z'}\gtrsim40$ MeV and coupling $g_{Z^\prime}\gtrsim6\times10^{-4}$, are the first search for a light $Z'$ (\emph{vanilla} $L_\mu-L_\tau$ model) with a muon beam using the missing energy-momentum technique (see Fig. \ref{fig:g-2-vanilla}).  
Figure \ref{fig:thermal-dm-limits} shows the obtained limits at 90\% CL in the target parameter space $(m_\chi,\ y)$ with freeze-out parameter $y=(g_\chi g_{Z'})^2(m_\chi/m_{Z'})^4$ for accelerator-based experiments probing thermal DM for $m_{Z'}=3m_{\chi}$, away from the resonant enhancement $m_{Z^\prime}\simeq2m_\chi$, and $g_\chi=5\times10^{-2}$. The thermal targets for favored $y$ values are plotted for scalar, pseudo-Dirac, and Majorana DM candidate scenarios, and obtained from the integration of the underlying Boltzmann equation \cite{Berlin:2018bsc}. The results indicate that NA64$\mu$ excludes a portion of the $(m_\chi,\ y)$ parameter space, below the current CCFR \cite{Altmannshofer:2014pba,CCFR:1991lpl} limits, constraining for a choice of masses $m_\chi\lesssim40$ MeV the dimensionless parameter to $y\lesssim6\times10^{-12}$.\\ \indent
In summary, for a total statistics of $(1.98\pm0.02)\times10^{10}$ MOT, no event falling within the expected signal region is observed.  Therefore, 90\% CL upper limits are set in the $(m_{Z^\prime},\ g_{Z^\prime})$ parameter space of the $L_\mu-L_\tau$ vanilla model, constraining viable mass values for the explanation of the $(g-2)_\mu$ anomaly to $6-7\ \text{MeV}\lesssim m_{Z'}\lesssim40$ MeV, with $g_{Z^\prime}\lesssim6\times10^{-4}$. New constraints on light thermal DM for values $y\gtrsim6\times10^{-12}$ for $m_\chi\gtrsim40$ MeV are also obtained. With improvements in the experimental set-up, such as an additional magnetic spectrometer to reduce by more than an order of magnitude  the background from momentum mis-recontruction,  and an increase in statistics, NA64$\mu$ is expected to fully cover the $(g-2)_\mu$ compatible parameter space and to boost its coverage in the search for thermal Dark Matter complementing the world wide effort for DS searches \cite{Jaeckel:2020dxj,Lanfranchi:2020crw,Krnjaic:2022ozp,Antel:2023hkf}. The use of a muon beam demonstrated in this work opens a new window to explore other well-motivated New Physics scenarios such as benchmark dark photon models in the mass region ($0.1-1$) GeV \cite{Gninenko:2019qiv}, scalar portals \cite{Sieber:2023nkq}, millicharged particles \cite{Gninenko:2018ter} or $\mu\rightarrow e$ or  $\mu\rightarrow \tau$ processes involving Lepton Flavour Conversion \cite{Gninenko:2018num,Gninenko:2022ttd,Radics:2023tkn}.\\ \indent
We gratefully acknowledge the support of the CERN management and staff, in particular the help of the CERN BE-EA department. We are grateful to C. Menezes Pires and R. Joosten for their support with the beam momentum stations. We are also thankful for the contributions from HISKP, University of Bonn (Germany), ETH Zurich, and SNSF Grants No. 186181, No. 186158, No. 197346, No. 216602, (Switzerland), ANID— Millennium Science Initiative Program—ICN2019 044 (Chile), RyC-030551-I and PID2021-123955NA-100 funded by MCIN/AEI/FEDER, UE (Spain). 
\bibliographystyle{apsrev4-1}
\bibliography{bibl}	

\end{document}